\RequirePackage[2020-02-02]{latexrelease}
\documentclass[aps,prl,twocolumn,showpacs]{revtex4}
\usepackage{amssymb}
\usepackage{amsmath}
\usepackage{graphicx}
\usepackage{color}

\begin{document}
\title{Crossover from inhomogeneous to homogeneous response \\of a  resonantly driven hBN quantum emitter}
\author{Domitille G\'erard, St\'ephanie Buil, Jean-Pierre Hermier, and Aymeric Delteil}

\affiliation{ Universit\'e Paris-Saclay, UVSQ, CNRS,  GEMaC, 78000, Versailles, France.\\ {\color{white}--------------------} aymeric.delteil@uvsq.fr{\color{white}--------------------}  }

%\date{\today }

\begin{abstract}
We experimentally investigate a solid-state quantum emitter ---a B center in hexagonal boron nitride (hBN)--- that has lifetime-limited coherence at short times, and experiences inhomogeneous broadening due to spectral diffusion at longer times. By making use of power broadening in resonant laser excitation, we explore the crossover between the inhomogeneous and the homogeneous broadening regimes. With the support of numerical simulations, we show that the lineshape, count rate, second-order correlations and long-time photon statistics evolve from a regime where they are dictated by spectral diffusion to a regime where they are simply given by the homogeneous response of the emitter, yielding restored Lorentzian shape and Poissonian photon statistics. Saturation of the count rate and line broadening occur not at the onset of the Rabi oscillations, but when the power-broadened homogeneous response becomes comparable with the inhomogeneous linewidth. Moreover, we identify specific signatures in both the second-order correlations and long-time photon statistics that are well explained by a microscopic spectral diffusion model based on discrete jumps at timescales of micro- to milliseconds. Our work provides an extensive description of the photophysics of B-centers under resonant excitation, and can be readily extended to a wide variety of solid-state quantum emitters.
\end{abstract}


 \maketitle
\section{I. Introduction}

While a multitude of physical systems have demonstrated the ability to emit single photons on demand, only a small subset are suitable for applications to quantum technologies, considering the highly stringent requirements set on the single-photon sources in terms of coherence, stability and count rate. In particular, single-photon emitters (SPEs) in the solid state have a significant potential for integration~\cite{Aharonovich16, Pelucchi22, Couteau23}, but typically suffer from the influence of the environment, which can yield pure dephasing~\cite{Cassabois08, Flagg09, Tighineanu18, White21} and spectral diffusion (SD)~\cite{Coolen09, Wolters13, Matthiesen14, Chen16, Spokoyny20, Fournier23PRB}. Pure dephasing is fundamentally detrimental to photon indistinguisability. On the other hand, even in the absence of such fast process ---if each of the emitted photons is perfectly coherent---, SD can cause effective dephasing by limiting the number of consecutive photons that can interfere with each other. However, it can in principle be counteracted by active feedback on the transition energy~\cite{Prechtel13, Hansom14}. It is therefore essential to establish appropriate characterisation methods to identify and characterize these processes in depth.

A combination of resonant laser excitation and photon correlations can efficiently characterize both dephasing and SD of individual quantum emitters~\cite{Fournier23PRB, Delteil24}. Based on this approach, in a prior work~\cite{Fournier23PRB}, we explored the purely inhomogeneous regime of a lifetime-limited emitter undergoing SD -- a B-center in hexagonal boron nitride (hBN). This family of emitters is characterized by a narrow electronic transition around 436~nm~\cite{Fournier21, Shevitski19, Gale22, Fournier23PRA, Horder22}. In the inhomogeneous broadening regime, if the atomic coherence is high enough, Rabi oscillations can nevertheless be observed in the $g^{(2)}$ function above emitter saturation, despite the time-averaged linewidth being given by a large inhomogeneous distribution. This is understood in a simple picture where only the close-to-zero laser-emitter detuning periods efficiently contribute to the correlation measurement based on two-photon coincidences, which prevents effective damping of the measured Rabi oscillations. This allows a direct extraction of the atomic coherence time. Spectral diffusion then impacts the second-order correlation function at longer times, since center wavelength fluctuations are converted to intensity fluctuations under cw resonant excitation~\cite{Fournier23PRB, Delteil24}.

Here, we take benefit from the power broadening phenomenon~\cite{Loudon} to use the laser power as a knob for tuning the spectral width of the homogeneous response of the emitter, allowing us to explore the crossover between the inhomogeneous and the homogeneous broadening regime of a resonantly driven emitter. The key is the observation of a single-photon emitter with a inhomogeneous linewidth a factor~10 above the natural linewidth, and that can be driven well above saturation, such that the power-broadened homogeneous linewidth exceeds the SD spectral distribution. This allows to explore the full range of regimes using a single quantum emitter.

Using a conceptually simple description, we consistently account for the entirety of the SPE photophysics under resonant excitation.  Our investigation includes the decay of Rabi oscillations, the line broadening, as well as the photon statistics at long times -- namely, the second-order correlation function and the intensity fluctuations of the photon counts. We show that the crossover is associated with an effective saturation of the count rate, which does not correspond to the onset of Rabi oscillations at $\Omega_R = \Gamma_1/\sqrt{2}$, but to the situation where the power-broadened homogeneous linewidth saturates the total linewidth, \textit{i.e.} $\Omega_R \sim \Delta \omega_\mathrm{inhom}$. 

Given that the particular microscopic process associated with SD can leave characteristic signatures in the photon statistics~\cite{Delteil24}, we then study the impact of SD on long-time photon correlations as well as the intensity fluctuations. Based on the observed signatures, we associate SD in B centers to a process that can be faithfully described by a Gaussian random jump model~\cite{Spokoyny20, Utzat19, Delteil24}, as we show based on statistical analysis and numerical simulations.

Our work provides a full picture of the photon statistics of spectrally diffusive two-level systems, applicable to both inhomogeneously broadened and close-to-homogeneous emitters. It sheds light on the apparent contradiction between saturation power deduced from Rabi oscillations and from count rate saturation. It also reconciles linewidth and coherence measurements based on Rabi oscillations in resonant excitation under arbitrary inhomogeneous broadening. Altogether, our investigation deepens the understanding of photon statistics of resonantly driven quantum emitters. By applying our approach to B-centers in hBN, we provide valuable additional knowledge about their photophysics and their spectral diffusion mechanisms.

\section{II. Experimental characterization of the emitter}
\label{II}

\subsection{A. Coherence time and inhomogeneous distribution}

The investigated physical system is a B center generated by electron irradiation~\cite{Fournier21, Gale22} in a high-pressure, high-temperature grown hBN crystal~\cite{Taniguchi07}. The characterization of the SPE is performed based on the methodology described in Ref.~\cite{Fournier23PRB}. The emitter is resonantly driven by a cw, tunable laser and the phonon sideband is collected and channeled to avalanche photodiodes.

\begin{figure}[h]
  \centering
  \includegraphics[width=0.7\linewidth]{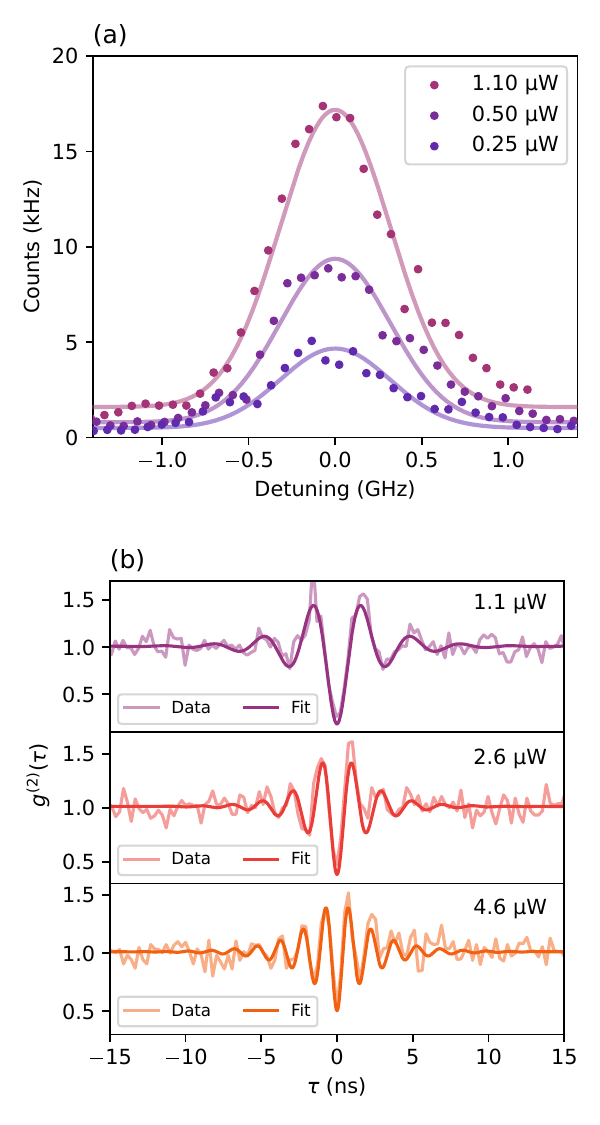}\\
  \caption{(a) Dots: Low power laser scans. Solid lines: Gaussian model. (b) Short-time $g^{(2)}(\tau)$. The solid lines are fits to the data using Eq.~\ref{shorttimeg2}.  }\label{lowPscan_Rabi}
\end{figure}

Figure~\ref{lowPscan_Rabi}a shows the results of three low-power laser scans. The emitter response exhibits a narrow peak that can be well described by a Gaussian function with a full width at half maximum of 0.75~GHz. This value, although narrow, is still about one order of magnitude larger than the natural linewidth ($\Gamma_1 = 82$~MHz) as deduced from the emitter lifetime $T_1 = \Gamma_1^{-1} =$~1.92~ns. To verify that this discrepancy is purely due to inhomogeneous broadening associated with fluctuations at timescales longer than the lifetime, we perform $g^{(2)}$ measurements. The results are shown Figure~\ref{lowPscan_Rabi}b for three different laser powers. Rabi oscillations can be observed and fitted with a model that fully accounts for the inhomogeneous broadening in the form:

\begin{equation}
\tilde{g}^{(2)}(\tau,\omega_L) = \dfrac{1}{N} \int d\omega S( \omega) C(\Omega_R,\omega_L, \omega)^2 g^{(2)}(\tau, \omega - \omega_L) 
 \label{shorttimeg2}
\end{equation}
with $S(\omega)$ the inhomogeneous distribution of the emitter center frequency $\omega$, $C(\Omega_R,\omega_L, \omega)$ the homogeneous response of the emitter, $g^{(2)}(\tau, \omega - \omega_L)$ the homogeneous second order correlation function, and $N$ a normalization constant. This expression generalizes equation~3 of ref.~\cite{Fournier23PRB}, valid only when $\Delta \omega_\mathrm{inhom} \gg \Delta \omega_\mathrm{hom}$, to any inhomogeneous distribution. It accounts for the fact that the measured correlations are the sum of the $g^{(2)}$ at all possible detunings weighted by their probability density, given by $S(\omega)$, and by the squared emitter population $C(\Omega_R,\omega_L, \omega)^2$, as should be the case for a two-photon coincidence measurement~\cite{Fournier23PRB, Koch23}. This narrowband weighting function ensures that only close-to-zero detuning periods contributes efficiently to the correlations, and therefore that the Rabi oscillations remain visible despite a potentially large inhomogeneous distribution. The data shown figure~\ref{lowPscan_Rabi}b are in excellent agreement with their fit by Eq.~\ref{shorttimeg2}, with $T_2 = 2 T_1$ at all powers (see Supplemental Information for the extended dataset). This observation confirms that the emitter response is lifetime-limited at short timescales, and therefore that the observed broadening is due to inhomogeneous broadening caused by slower processes. 

The mechanism for inhomogeneous linewidth broadening is attributed to fast spectral diffusion occuring at timescales from micro- to milliseconds, as confirmed and further analyzed in section~III. The fits of Rabi oscillations also provide a direct correspondence between laser power and Rabi frequency (see Supplemental Information). This allows us to infer the power $P_\mathrm{sat} \approx $~40~nW associated with saturation of the two-level system, which is such that $\Omega_R = \Gamma_1/\sqrt{2}$.

Based on the previous observations, in the following, we describe the SPE as a maximally coherent emitter having a power-independent Gaussian inhomogeneous distribution of the center wavelength $S( \omega) = 1/\sqrt{2\pi}\Sigma \exp(-\omega^2/2\Sigma^2)$, with a full width at half maximum (FWHM) $\Delta\omega_\mathrm{inhom} = 0.75$~GHz, and a homogeneous linewidth at low power given by $\Delta\omega_\mathrm{hom} = \Gamma_1 = 82$~MHz.

\subsection{B. Effective saturation of the spectrally diffusing emitter}

The average count rate under resonant excitation with a laser frequency $\omega_L/2\pi$ is given by:

\begin{align}
\bar{C}(\Omega_R, \omega_L) &= \langle C(\Omega_R, \omega_L, \omega)\rangle_\omega \nonumber \\
&=  \int_{-\infty}^{+\infty} S(\omega) C(\Omega_R, \omega_L, \omega)d\omega \nonumber \\
&=  \dfrac{1}{4\Sigma\sqrt{2 \pi}} \int_{-\infty}^{+\infty} 
e^{-\dfrac{\omega^2}{2 \Sigma^2}}
\dfrac{\Omega_R^2}{\omega^2 + \dfrac{\Gamma_1^2}{4} +\dfrac{ \Omega_R^2}{2}}
d\omega
\label{lineshape}
\end{align}

Since the increase of the laser power results in an increase of the linewidth of $C(\Omega_R, \omega_L, \omega)$, it allows to tune the ratio $\Delta\omega_\mathrm{hom}/\Delta\omega_\mathrm{inhom}$. Such increase has consequences on both the count rate and the lineshape.

\begin{figure}[h]
  \centering
  \includegraphics[width=0.95\linewidth]{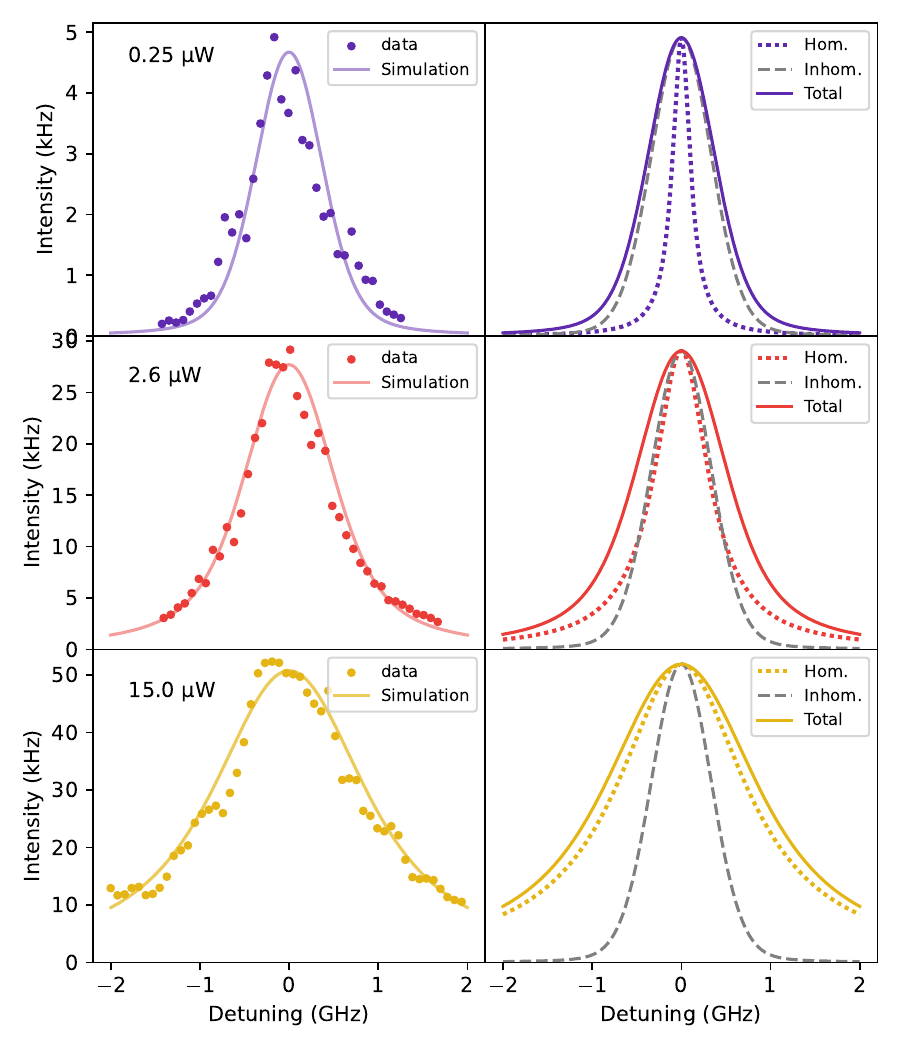}\\
  \caption{Left column: simulated (plain lines) and measured (dots) count rate as a function of the laser detuning, at three different powers. Right columns: Calculated homogeneous (dotted line), inhomogeneous (dashed line) and total (plain line) contributions to the lineshape.}\label{Broadening}
\end{figure}

Concerning the lineshape, when $\Delta\omega_\mathrm{inhom}$ becomes comparable to $\Delta\omega_\mathrm{hom}$, the time-average response starts to broaden and the homogeneous response dominates the lineshape. For Gaussian inhomogeneous distribution, this is well known to give rise to a Voigt profile, that evolves towards a purely Lorentzian lineshape. Figure~\ref{Broadening} displays experimental laser scans (dots) together with the results of Eq.~\ref{lineshape} (plain lines) at three different powers chosen in the three different regimes (\textit{i.e.} inhomogeneous, intermediate, and homogeneous), with $\Omega_R$, $\Delta \omega_\mathrm{inhom}$ and $\Gamma_1$ fixed to the values extracted from the data of the previous section. The calculated lineshapes are in good agreement with the data. This is also an additional confirmation that $T_2 = 2 T_1$, since at high power the linewidth is approximately given by $2 \Omega_R \sqrt{T_1/T_2}$. Therefore, the presence of sizable pure dephasing would have yielded broader lines than we both measured and calculated at high powers.

\begin{figure}[t]
  \centering
  \includegraphics[width=\linewidth]{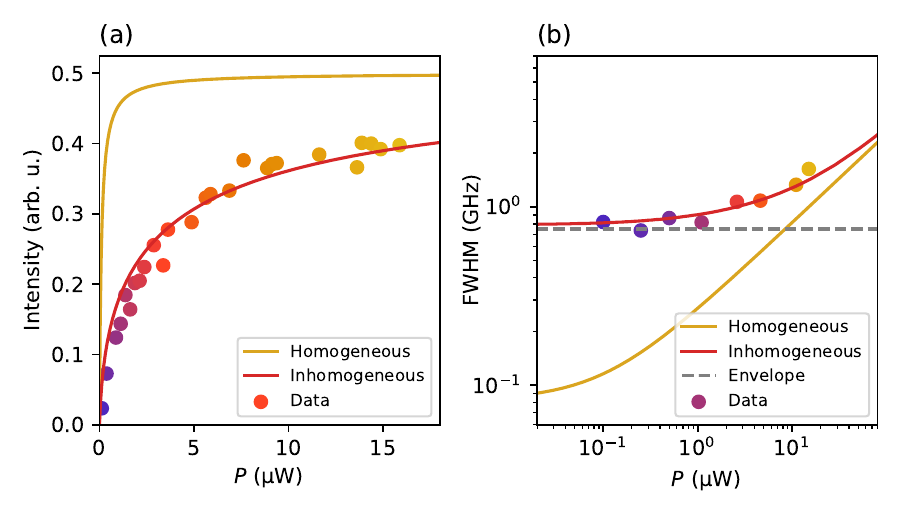}\\
  \caption{(a) Dots: Count rate as a function of the laser power. Red line: Model of Eq.~\ref{eq_saturation}. Yellow line: Saturation curve in the absence of spectral diffusion. (b) Dots: Measured linewidth as a function of the laser power. Red line: Linewidth calculated from Eq.~\ref{lineshape}. Yellow line: Linewidth in the absence of spectral diffusion.}\label{Psat_lw}
\end{figure}

We now come to the dependence of the maximum count rate on the laser power. The dots on Figure~\ref{Psat_lw}a show the count rate measured with the laser tuned to resonance ($\omega_L = 0$) as a function of the laser power. It exhibits a saturation behavior similar to that of a two-level system in the absence of spectral diffusion. However, the saturation occurs at a power ($\sim 2.5$~$\mu$W) well above the saturation power calculated in section II.A from the Rabi oscillations. This can be understood in the picture of a spectrally diffusive emitter. Indeed, when driven on resonance, even when the homogeneous response has saturated, the count rate continues to grow due to the increasing overlap between the laser and the homogeneous response, which is a consequence of power broadening. This is captured by Eq.~\ref{lineshape}, which leads to the following analytical expression when the laser is tuned at the center of the inhomogeneous distribution:

\begin{equation}
\bar{C}(\Omega_R, \omega_L = \omega_0) = \left(\dfrac{\Omega_R}{\Delta \omega_\mathrm{hom}} \right)^2 f \left(  
\dfrac{\Delta \omega_\mathrm{hom}}{2\sqrt{2} \Sigma}
 \right)
\label{eq_saturation}
\end{equation}

with 

\begin{equation}
f(x) = \sqrt{\pi}x \exp\left(x^2\right) \mathrm{erfc}(x)
\label{eq_f}
\end{equation}

The first contribution in Eq.~\ref{eq_saturation} describes the saturation of the two-level system and rules the behavior in the absence of SD (where $f(x) = 1$). In the general case, when $\Gamma_1 < \Omega_R < \Sigma$ -- \textit{i.e.} above $P_\mathrm{sat}$ but below the homogeneous regime, the term $f (\Delta \omega_\mathrm{hom}/2\sqrt{2} \Sigma )$ describes the increase of count rate due to power broadening. The function saturates when $\Omega_R \gtrsim \Sigma$ since the saturated homogeneous line dominates the response, and SD has then vanishing effect on the count rate.

The result of this equation is plotted on figure~\ref{Psat_lw}a (red plain line). It exhibits a qualitatively similar saturation behavior as a non-diffusing two-level system, albeit saturates at a higher power given by $P_\mathrm{sat}^\mathrm{eff} \approx 3/2\left( \Sigma/\Gamma_1 \right)^2 P_\mathrm{sat}$. This effective saturation does not take place at the onset of the Rabi oscillations (\textit{i.e.} $P\sim P_\mathrm{sat}$) but at a power such that the probability the emitter is on resonance with the laser saturates. This occurs when the homogeneous linewidth becomes the dominant contribution to the total linewidth. This effective saturation power therefore marks the transition from the inhomogeneous to the homogeneous response of the emitter. In our case, $P_\mathrm{sat}^\mathrm{eff}$ is about 2.5~$\mu$W, almost two orders of magnitude higher than $P_\mathrm{sat}$, which is consistent with the values of $\Sigma$ and $P_\mathrm{sat}$ established in section~II.A. We also plot the corresponding theoretical saturation curve in the absence of spectral diffusion, given by $C \propto 1/(1+P_\mathrm{sat}/P)$. It should be noted that the asymptotic population is always equal to 1/2, regardless of the presence of SD, since the high-power response is always homogeneous -- therefore, the asymptotic count rate is independent of SD.

The considerations developed in this section are independent of the particular spectral diffusion mechanism and timescale -- provided that it is sizeably longer than the lifetime, which is guaranteed by the observation of maximally coherent Rabi oscillations at short times. Our conclusions are also, to a large extent, independent of the particular lineshape of the inhomogeneous distribution.

\section{III. Long-time photon statistics and spectral diffusion model}
\label{III}

\subsection{A. Bunching in the long-time $g^{(2)}$}

\begin{figure}[t]
  \centering
  \includegraphics[width=0.95\linewidth]{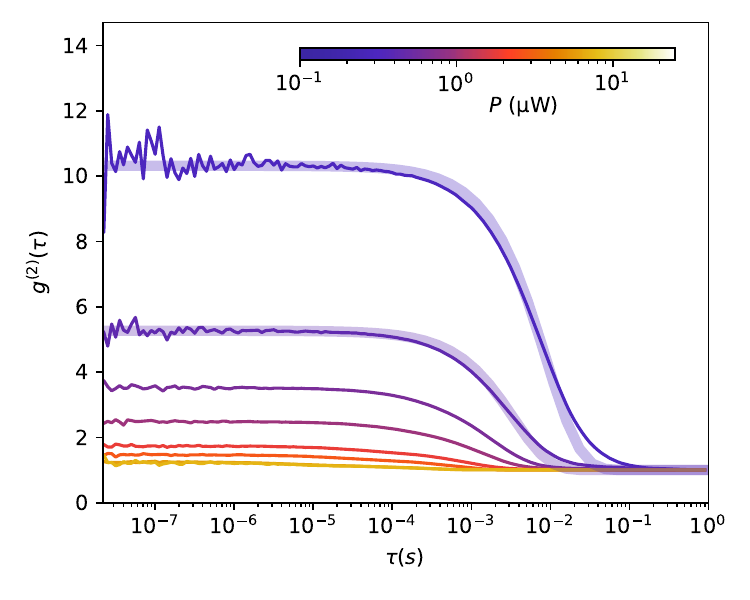}\\
  \caption{Thin lines: $g^{(2)}$ function measured at different powers. Thick lines: Monoexponential fit to the data.}\label{LongTimeG2}
\end{figure}

As established in prior work~\cite{Fournier23PRB, Delteil24}, a resonantly driven, spectrally diffusive quantum emitter exhibits power-dependent bunching at timescales longer than the lifetime. This is also the case of the emitter investigated here, as can be seen on figure~\ref{LongTimeG2} where we plot the $g^{(2)}$ measured at different powers over height orders of magnitude of delay times. At low power, the $g^{(2)}$ starts with a high value (of the order of $\Delta\omega_\mathrm{inhom}/\Delta\omega_\mathrm{hom}$) at short times, then decays at a timescale governed by SD. At higher powers, the height of the short-time plateau decreases due to power broadening. We note that, at the highest powers, the amount of short-time bunching is vanishing, consistently with the picture of a homogeneous response of the resonantly driven emitter. The decay shape of the normalized bunching $\tilde{B}(\tau) = (g^{(2)}(\tau) - 1)/(g^{(2)}(0) - 1)$ is characteristic of the associated microscopic diffusion process~\cite{Delteil24}. In the present case, the bunching decay can be fitted with an exponential decay (thick lines on figure~\ref{LongTimeG2}), which is associated with a Gaussian random jump (GRJ) process. This process consists in discrete Poissonian events that make the emitter center frequency jump to a new value given by the Gaussian distribution that defines the inhomogeneous lineshape~\cite{Spokoyny20, Utzat19, Delteil24}. It can be seen on figure~\ref{LongTimeG2} that the monoexponential fit is not perfect, and a finer analysis actually reveals two very close timescales that split apart at increasing power (see Supplemental Material). In the following, for simplicity we make the approximation of a single GRJ process with a characteristic time $\tau_\mathrm{SD}$ given by the single-exponential fit parameter.

\subsection{B. Time traces and count rate histograms}

\begin{figure}[t]
  \centering
  \includegraphics[width=0.95\linewidth]{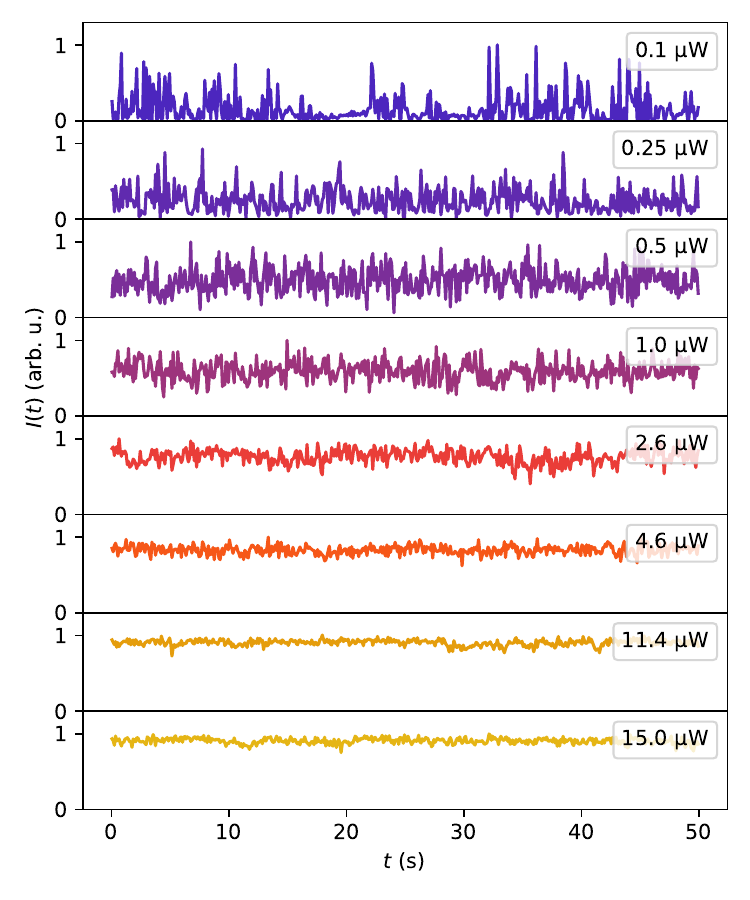}\\
  \caption{Intensity time traces measured at eight different powers, with a 100~ms binning.}\label{ExpTimetraces}
\end{figure}

It has been predicted that a discrete SD process such as the GRJ model yields qualitatively different intensity fluctuations as compared with a continuous diffusion process~\cite{Delteil24}. In order to confirm that the GRJ model is a valid candidate for a microscopic description of the SD process occuring for B centers, we examine the intensity-time traces at various laser powers, at zero laser-emitter average detuning. Figure~\ref{ExpTimetraces} displays the count rate as a function of time using a 100~ms binning time, for various powers spanning the three regimes. It can be seen that the intensity fluctuations strongly depend on the power. In particular, at low powers, the count rate exhibits sequences of short bright periods separated by longer dark periods~\cite{Horder24}. When the power then increases, the fluctuations becomes more regular, with a stable count rate.

\begin{figure}[!b]
  %\centering
  \includegraphics[width=\linewidth]{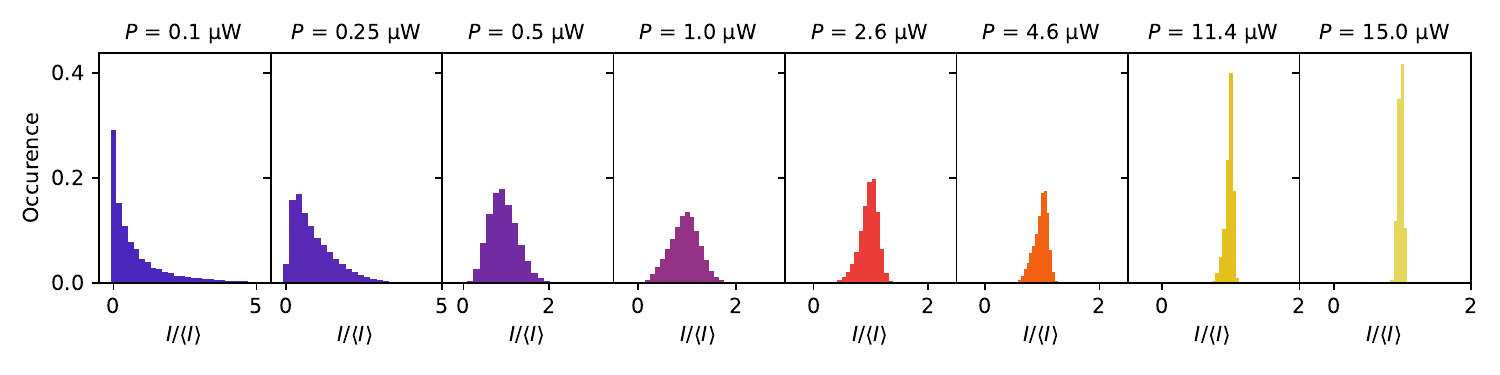}\\
  \caption{Intensity histograms corresponding to the eight time traces shown Fig.~\ref{ExpTimetraces}.}\label{ExpHists}
\end{figure}

To obtain deeper insights on the statistics of these intensity fluctuations, on Figure~\ref{ExpHists} we plot the corresponding intensity histograms. At low powers ($P \ll P_\mathrm{sat}^\mathrm{eff}$), the histograms exhibit a positive skewness. It originates from the fact that most of the time, the emitter is out of resonance with the laser, yet during the rare periods where it is close to resonance with the laser, it yields short but bright intensity bursts. When increasing the power, the histogram shape evolves from a positive skewness to a negative skewness at higher powers ($P \sim P_\mathrm{sat}^\mathrm{eff}$). This is a feature of the onset of the homogeneous regime. Indeed, due to power broadening, at powers close to $P_\mathrm{sat}^\mathrm{eff}$, the emitter turns out to spend most of its time close to resonance. The intensity is therefore more stable, but occasionally larger frequency fluctuations can decrease the count rate, giving rise to the low-intensity tail. Finally, above effective saturation ($P \gg P_\mathrm{sat}^\mathrm{eff}$), the response is mostly homogeneous and SD has negligible effect on the count rate. The intensity histogram therefore narrows down, the skewness vanishes and the count rate becomes stable. The noise is then dominated by Poisson noise characteristic of a homogeneous regime. This is consistent with the measured $g^{(2)}(\tau)$ shown figure~\ref{LongTimeG2}, which tends to a coherent response, and hence to purely Poissonian fluctuations associated with a vanishing Mandel parameter~\cite{Delteil24}.

\onecolumngrid
\vspace*{0.8cm} 
\twocolumngrid  % Switch back to two-column mode
~ \\
~ \\

\begin{figure}[t]
  \centering
  \includegraphics[width=\linewidth]{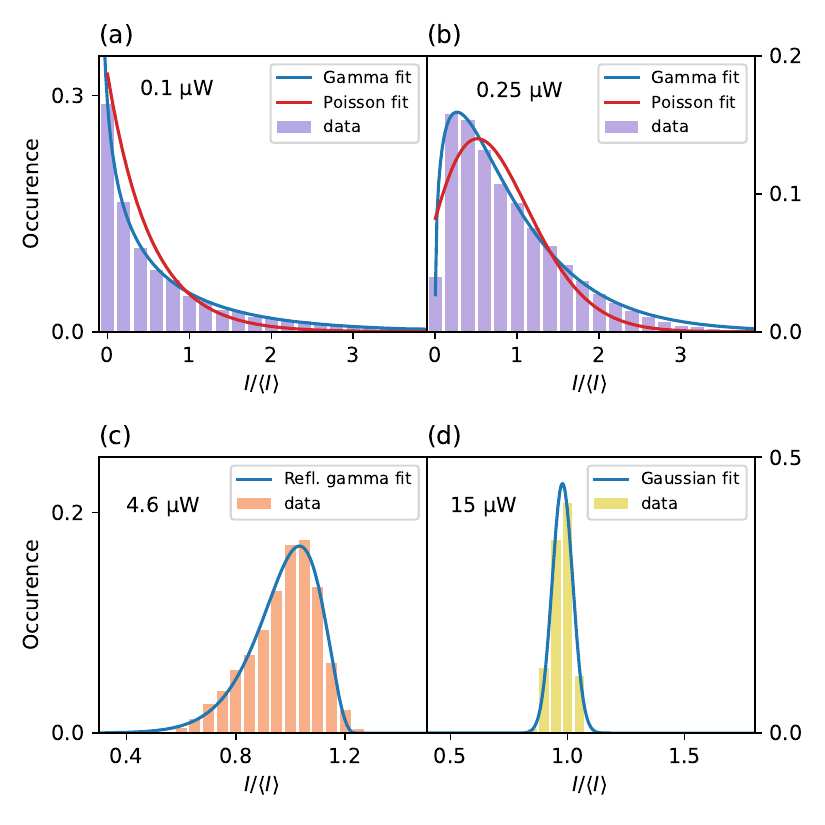}\\
  \caption{(a) Bars: Low-power (0.1~$\mu$W) intensity histogram. Blue curve: Fit with a gamma distribution. Red curve: Fit with a Poisson distribution. (b) Same as (a) at power 0.25~$\mu$W. (c) Bars: Medium-power (4.6~$\mu$W) intensity histogram. Blue curve: Reflected gamma fit to the histogram.(d) Bars: High-power (15~$\mu$W) intensity histogram. Blue curve: Gaussian fit to the histogram.}\label{HistFits}
\end{figure}

 This behavior can be quantitatively understood in the frame of the GRJ model. In the low power limit, the intensity histogram shape has been predicted to follow a gamma distribution~\cite{Delteil24}, owing to the exponential distribution of bright period durations. This is opposite to continuous diffusion models, which yield Poissonian intensity distributions even at low powers. Figures~\ref{HistFits}a and b show a fit to the two lowest-power histograms with both a gamma and a Poisson distribution. It can be observed that the data are very well fitted by a gamma distribution, while the agreement with a Poisson distribution is much poorer. This confirms the hypothesis that the GRJ model faithfully describes SD in hBN B-centers. By a symmetric reasoning, it is conceivable that the intermediate regime histograms should follow a reflected gamma distribution, due to a generally close-to-maximum intensity interspersed with darker periods whose duration follows an exponential distribution. We find that the histograms associated with this regime is indeed well fitted by a reflected gamma distribution, as can be seen on Figure~\ref{HistFits}c. Finally, the high-power histogram can be fitted using a symmetric and narrow Gaussian distribution, as visible on Figure~\ref{HistFits}d.

\begin{figure}[t]
  \centering
  \includegraphics[width=\linewidth]{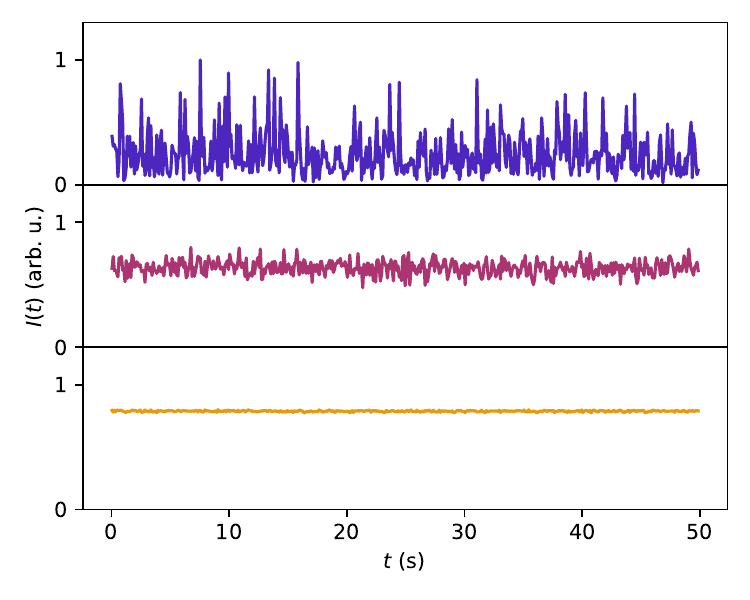}\\
  \caption{Intensity traces simulated using the GRJ model at different powers.}\label{TimeTracesSimu}
\end{figure}

As a final confirmation for the jump-based mechanism, we simulate time traces using a GRJ model into which we have entered our experimental parameters (\textit{i.e.} $\Sigma$, $\tau_\mathrm{SD}$ and $\Gamma_1$). Figure~\ref{TimeTracesSimu} shows a few examples of time traces we have generated, using the same binning as in figure~\ref{ExpHists}. A good qualitative agreement can be observed between the simulated and the experimental time traces. In particular, the low-power fluctuations in form of intensity spikes, or bursts, are well reproduced by the model and confirm that jump-like SD governs the photon statistics of the resonantly driven B center. 

\section{IV. Conclusion}

Our work consolidates and generalizes the description of photon statistics of resonantly driven quantum emitters subject to spectral diffusion, and illustrates the variety of statistical properties of photon detection events in the whole range of regimes that spans from the purely inhomogeneous to the homogeneous response. Furthermore, our investigation deepens the understanding of B centers in hBN and allows to establish the GRJ model as an accurate description of their inhomogeneous broadening mechanism. 

\section{Acknowledgments}
The authors thank Clarisse Fournier for fruitful discussions. This work is supported by the French Agence Nationale de la Recherche (ANR) under reference ANR-21-CE47-0004-01 (E$-$SCAPE project).

\end{document}